\documentclass[10pt,conference]{IEEEtran}
\IEEEoverridecommandlockouts

\usepackage[T1]{fontenc}
\usepackage[utf8]{inputenc}
\usepackage{microtype}
\usepackage{xspace}
\usepackage[inline]{enumitem}

\usepackage{amsmath}
\usepackage{amssymb}
\usepackage{amsthm}
\usepackage{stmaryrd}

\usepackage{graphicx}
\usepackage{booktabs}
\usepackage{tabularx}
\usepackage{multirow}
\usepackage{xcolor}
\usepackage[most]{tcolorbox}
\usepackage{framed}
\usepackage{pifont}
\usepackage{subcaption}
\usepackage{tikz}
\usetikzlibrary{arrows.meta,positioning,fit,backgrounds,calc}

\usepackage{listings}
\usepackage[ruled,vlined,linesnumbered]{algorithm2e}

\usepackage{mathpartir}

\usepackage[hidelinks]{hyperref}
\usepackage{url}
\usepackage[capitalize,noabbrev]{cleveref}
\crefname{observation}{Observation}{Observations}
\Crefname{observation}{Observation}{Observations}

\theoremstyle{plain}

\theoremstyle{definition}
\newtheorem{definition}{Definition}

\newcommand{\code}[1]{\texttt{#1}}
\newcounter{rq}

\definecolor{formalshade}{rgb}{0.95,0.95,0.97}
\definecolor{darkblue}{rgb}{0.14,0.22,0.52}
\newenvironment{takeaway}{
\small

\MakeFramed{\advance\hsize-\width\FrameRestore}}
{\endMakeFramed}

\newcommand{\rules}{\mathcal{R}}

\newcommand{\Reify}{\textsc{Reify}}
\newcommand{\Disperse}{\textsc{Disperse}}
\newcommand{\Offload}{\textsc{Offload}}
\newcommand{\attack}{\textsc{SkillCloak}\xspace}
\newcommand{\defense}{\textsc{SkillDetonate}\xspace}

\begin{document}

\title{Cloak and Detonate: Scanner Evasion and Dynamic Detection of Agent Skill Malware}

\author{
\IEEEauthorblockN{
Zimo Ji\textsuperscript{1},
Congying Xu\textsuperscript{1,2,*},
Zongjie Li\textsuperscript{1},
Yudong Gao\textsuperscript{1},
Xin Wei\textsuperscript{1},
Shuai Wang\textsuperscript{1,*},
Shing-Chi Cheung\textsuperscript{1,2}
}
\IEEEauthorblockA{
\textsuperscript{1}Hong Kong University of Science and Technology\quad
\textsuperscript{2}Guangzhou HKUST Fok Ying Tung Research Institute, China\\
\{zjiag, zligo, shuaiw, scc\}@cse.ust.hk,\quad
\{congying.xu, ygaodj, xweiba\}@connect.ust.hk\\
\textsuperscript{*}Corresponding authors
}
}

\maketitle

\begin{abstract}

LLM coding agents increasingly rely on third-party \emph{agent skills} from public marketplaces, which execute with the agent's privileges and create a software supply-chain attack surface: a malicious skill can steal credentials, exfiltrate source code, or install backdoors. Existing defenses use static skill scanners based on pattern matching or LLM-as-judge analysis, but it remains unclear whether they withstand adaptive evasions that preserve malicious behavior while changing payload appearance.

This paper first presents an adversarial study of existing skill scanners through \attack, a payload-preserving evasion framework that keeps the attack semantics intact while transforming their visible form. \attack uses two complementary strategies: \emph{Structural Obfuscation}, which rewrites visible payload indicators into semantically equivalent forms, and \emph{Self-Extracting Skill (SFS) Packing}, which hides malicious components from the install-time view and restores them during agent execution. Across eight scanners and 1{,}613 in-the-wild malicious skills, \emph{SFS Packing} bypasses every scanner at over 90\%, while \emph{Structural Obfuscation} bypasses over 80\% on most static scanners and reaches 96\% on a hybrid scanner, showing that appearance-based auditing is insufficient.

Motivated by this finding, we propose \defense, a behavior-centric runtime auditor that executes skills in a sandbox and detects malicious effects through OS-boundary information-flow evidence rather than install-time appearance. \defense combines on-demand closure lift, which observes instructions materialized during execution, with marker-based taint analysis, which tracks sensitive-data flows across the agent context, files, processes, and network operations. The results show that \defense detects 97\% of attacks at a 2\% false-positive rate and sustains 87\% detection on real-world malicious skills.

\end{abstract}


\section{Introduction}
\label{sec:intro}

LLM agents are rapidly moving from demonstrations to real-world deployment across domains, including software engineering (e.g., ``vibe coding’’ assistants~\cite{jimenez2024swebench}), cybersecurity~\cite{zhang2025cybench}, and autonomous driving~\cite{mao2023gptdriver}.
A key enabler of this transition is the emergence of \emph{Agent Skills}: modular packages that allow agents such as Claude Code~\cite{anthropic2026claudecode} and OpenAI Codex~\cite{openai2026codex} to acquire new capabilities on demand.
A skill is typically distributed as a structured directory containing natural-language instructions (e.g., \code{SKILL.md}), executable scripts or code blocks, and auxiliary resources that an agent can load and invoke during task execution~\cite{anthropic2025skills}.
Because skills are model-agnostic, composable, and shareable as ordinary files, they are increasingly becoming a unit of capability distribution for agentic systems, with public skill marketplaces growing rapidly~\cite{anthropic2025skills, marketsandmarkets2026agents}.
The scale of this ecosystem is already substantial: within months of the open standard being introduced in late 2025, a single marketplace had accumulated over 40{,}000 publicly listed skills~\cite{skills2026analysis}, the overwhelming majority of which are community-contributed and unvetted.

This extensibility, however, introduces a new software supply-chain risk.
Once installed, a skill is interpreted and executed by an agent that may have access to the developer’s workspace, local files, credentials, package managers, terminals, and external services.
A malicious skill can therefore abuse these inherited privileges to perform security-sensitive actions, including credential theft, source-code exfiltration, backdoor installation, and destructive file operations.
For instance, the recent \emph{ClawHavoc} campaign planted over 300 malicious skills on a single public marketplace, whose \code{SKILL.md} files instructed the agent to fetch and run an information stealer under the guise of an installation prerequisite, silently harvesting the victim's browser credentials, keychain passwords, SSH keys, and cryptocurrency wallets~\cite{koi2026clawhavoc}.
The attack surface is particularly broad, and the malicious payload can be embedded in the natural-language instructions of \code{SKILL.md}, or hidden in bundled scripts or resources.
A range of malicious behaviors have already been observed in published skills, including credential theft~\cite{chen2026credential,koi2026clawhavoc}, data exfiltration~\cite{snyk2026toxicskills,liu2026skillswild}, and backdoor deployment~\cite{malskills2026wild}.

\noindent \textbf{Problem.}
To mitigate these risks, recent efforts have proposed skill scanners that audit third-party skills before installation~\cite{skillsieve2026,cascade2026,malskills2026}.
They are designed to recognize suspicious patterns in the skill’s visible files, using static rules, LLM-based judgments, or their combination.
However, it remains unclear whether such static auditing mechanisms can withstand adaptive evasion.
An adversarial skill author is not constrained to expose the malicious payload in the form expected by the scanner; the payload can be rewritten, relocated, or or staging the payload until runtime. This raises a fundamental question: \emph{To what extent can current skill auditing mechanisms detect malicious skills under semantics-preserving evasive transformations?}

\begin{figure}[t]
  \centering
  \includegraphics[width=.95\columnwidth]{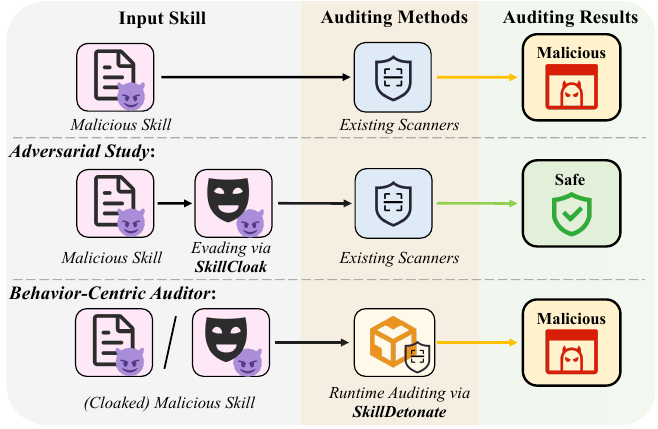}
  \caption{Overview of \attack and \defense.}
  \label{fig:overview}
  \vspace{-10pt}
\end{figure}

As \Cref{fig:overview} illustrates, static scanners flag a malicious skill, but
\attack rewrites it into a variant they clear as \emph{safe}, whereas \defense
detects both the original and cloaked skills by auditing their runtime behavior.

\noindent \textbf{Adversarial Study.}
To address this problem, we first conduct an adversarial study on eight representative skill scanners and two production agents (Codex and Claude Code) with 1{,}613 \emph{in-the-wild} malicious skills.
We propose \attack, a payload-preserving evasion framework that adaptively transforms malicious skills while preserving their malicious behavior.

\attack operates on two complementary strategies (Section~\ref{sec:skillcloak}).
\begin{enumerate}
  \item \emph{Structural Obfuscation} \emph{rewrites} concrete payload indicators (e.g., malicious instructions, shell commands, URLs, and credential-related strings) into semantically equivalent forms that evade static rules and LLM-based judgments. This strategy perturbs the skill's structure minimally, making it stealthy with strong bypass effectiveness.
  \item \emph{Self-Extracting Skill Packing (SFS Packing)} \emph{hides} malicious instructions or scripts outside the scanner's observable scope (e.g., ignored directories or encoded resource blobs) and restores them at execution time. This strategy alters the skill's structure substantially, but achieves near-complete bypass effectiveness.
\end{enumerate}

We collected 1{,}613 \emph{in-the-wild} malicious skills from the ClawHub skills archive~\cite{clawhub2026archive}, and employ \attack to generate evasion skills for adversarial evaluation.
The results show that
\begin{enumerate}
  \item existing skill scanners are highly vulnerable to semantics-preserving evasions: with \emph{SFS Packing}, over $90\%$ of evasion skills bypass all scanners, while \emph{Structural Obfuscation} bypasses over $80\%$ on most static scanners and up to $96\%$ on a hybrid scanner.
  \item \attack generates payload-preserving evasion skills that stay functional: across both Codex and Claude Code, cloaking causes no statistically detectable loss in skill utility.
\end{enumerate}
These findings show \textit{the limits of existing static skill auditing against evasive malware and motivate the need of behavior-centric defense}.

\noindent \textbf{Behavior-Centric Auditor.}
To meet this need, we propose \defense, a behavior-centric runtime auditor for agent skills.
The key \textit{insight} is that evasion can alter how a payload is written, packaged, or staged, but they must produce observable behaviors and effects to achieve a malicious objective.
\defense therefore executes a suspicious skill in a controlled sandbox and monitors the security-relevant effects it produces during agent execution.

To achieve this, \defense introduces two novel designs.
\begin{enumerate}
  \item \emph{On-Demand Closure Lift} captures and executes the
  natural-language instructions a skill materializes at runtime, feeding them
  back into the agent's session. This mitigates the path-coverage gaps typically
  exploited by multi-stage malware.
  \item \emph{Marker-Based Taint Analysis} tracks information flow across the two media a syscall monitor cannot natively follow. Data markers, planted at sensitive reads intercepted through a FUSE virtual filesystem, carry provenance through the agent's natural-language context; and inode-level reconstruction of the eBPF syscall graph recovers that same flow across opaque cross-process byte transforms.
\end{enumerate}

We evaluated \defense on SkillJect~\cite{skillject2026}, a benchmark of $150$ malicious skills that inject prompt-injection payloads into skill-enabled agents.
The results show that \defense can successfully detect $97\%$ of malicious skills at a $2\%$ false-positive rate, $31\%$ more than the best static scanner and $62\%$ more than a naive OpenCode based agent detector.
On real-world malicious skills from MalSkillBench~\cite{malskillbench2026}, \defense sustains $87\%$ detection and stays stable under \attack, whereas the best static scanner collapses from $99\%$ to $10\%$ under Structural Obfuscation.
In this paper, we make the following contributions.

\begin{itemize}
	\item To the best of our knowledge, we are the \textbf{first to evaluate existing skill auditing mechanisms against adaptive semantics-preserving evasion}.
The results show that existing auditing mechanisms are highly vulnerable to semantics-preserving evasions.
	\item We propose \attack, \textbf{a skill evasion framework} to transforms malicious skills while preserving the payload via two complementary strategy (\textit{Structural Obfuscation} and \textit{Self-Extracting Skill Packing}).
	\item We propose \defense, \textbf{a behavior-centric runtime auditor} for agent skills to detect malicious skills by observing security-relevant effects during agent execution with two novel designs (\textit{On-Demand Closure Lift} and \textit{Marker-Based Taint Analysis}).
	\item We conduct extensive experiments to evaluate the effectiveness of \defense.
\end{itemize}

\section{Background and Goal}
\label{sec:background}

\subsection{Malicious Agent Skills.}
An agent skill is a structured directory bundle containing natural-language instructions, executable components, and auxiliary resources that a host agent discovers by name and loads into its context to gain a capability~\cite{anthropic2025skills}.
These contents are typically organized as a \code{SKILL.md} instruction file, optional \code{scripts/} or code blocks, and optional \code{references/} or data files.
Because the agent-loaded context is treated as trusted instructions and its scripts run with the agent's privileges, a poisoned bundle is a direct vehicle for malicious behavior on a coding agent~\cite{malskillbench2026,malskills2026wild}.

Moreover, the adversary need not concentrate the payload in one place: a natural-language directive in \code{SKILL.md} instructs the agent to run a bundled script, while the script itself carries the exfiltration logic, so the malicious intent is split across the two media a skill ships.
Despite this exposed surface, offensive research on skills remains thin and largely
\emph{benchmark-} or \emph{template-level}: existing work measures how often
agents follow injected instructions, or supplies hand-written poisoned
templates~\cite{skillject2026,schmotz2026skillinject,qu2026supplychain}.

What is missing is a systematic, software-and-systems account of how an
adversary can engineer a skill to defeat the deployed defenses while keeping its
payload intact. \attack (\Cref{sec:skillcloak}) fills this gap.

\subsection{Agent Skill Auditing.}
\label{sec:background:auditing}
The defensive response has centered on skill scanners: tools that vet a bundle at install time and emit a block/allow verdict for continuous integration (CI) gating.
We surveyed nine representative scanners (\Cref{tab:scanners}). Most reduce to
two mechanisms, often combined. \emph{Static pattern matching} applies regex,
YARA, or AST rules over the file tree: Huifer~\cite{scanner_huifer} and
Vigile~\cite{scanner_vigile} run regex rulebooks (a YAML rule set and 59
TypeScript patterns), Qualixar~\cite{scanner_qualixar} infers a skill's
capabilities by AST and flags those exceeding its declared permissions,
Pors~\cite{scanner_pors} orchestrates off-the-shelf tools (shellcheck,
trufflehog, and semgrep), and Headyzhang~\cite{scanner_headyzhang} adds
tool-boundary AST taint tracking; none of these four invoke a model.
\emph{LLM-as-judge} instead serializes the bundle into a prompt and asks a
model for a verdict, as in Patidarganesh~\cite{scanner_patidarganesh}, which
dumps the readable files against a fixed threat checklist. The rest combine the
two: Caterpillar~\cite{scanner_caterpillar} and Cisco~\cite{scanner_cisco} run
static rules (regex, YARA signatures, and dataflow taint) and then gate the
result through an optional LLM stage, while Nova~\cite{scanner_nova} stacks
keyword, semantic, and LLM matching. The first two columns of
\Cref{tab:scanners} summarize each scanner and its matching pattern.

\begin{table}[t]
\caption{Overview of the nine surveyed skill scanners. \textbf{Par.}:
\textbf{S}=static, \textbf{H}=hybrid, \textbf{J}=LLM-judge. \emph{Blind-dir} /
\emph{Blind-file}: a directory / file the scanner never reads.
}
\label{tab:scanners}
\centering
\footnotesize
\setlength{\tabcolsep}{3.5pt}
\begin{tabular}{@{}l c l l l@{}}
\toprule
Scanner & Par. & Pattern & Blind-dir & Blind-file \\
\midrule
Huifer~\cite{scanner_huifer}             & S & regex     & \texttt{build/} & $\notin\{$\texttt{.md,.py,…}$\}$ \\
Caterpillar~\cite{scanner_caterpillar}   & H & regex+LLM & \texttt{.git/}  & $\notin\{$\texttt{.md,.py,…}$\}$ \\
Headyzhang~\cite{scanner_headyzhang}     & S & AST       & \texttt{.git/}  & $\notin\{$\texttt{.py,.ts,…}$\}$ \\
Vigile~\cite{scanner_vigile}             & S & regex     & \texttt{.git/}  & $\neq$\,\texttt{SKILL.md} \\
Qualixar~\cite{scanner_qualixar}         & S & AST       & \texttt{.git/}  & $\neq$\,\texttt{SKILL.md} \\
Pors~\cite{scanner_pors}                 & S & SAST      & \texttt{-}      & $\notin\{$\texttt{.md,.sh,…}$\}$ \\
Cisco~\cite{scanner_cisco}               & H & YARA+LLM  & \texttt{.git/}  & binary \\
Nova~\cite{scanner_nova}                 & H & sem+LLM   & \texttt{docs/}  & $\notin\{$\texttt{.md,.txt,…}$\}$ \\
\midrule
Patidarganesh~\cite{scanner_patidarganesh} & J & LLM     & \texttt{.git/}  & $\notin\{$\texttt{.md,.py,…}$\}$ \\
\bottomrule
\end{tabular}
\end{table}

\subsection{Threat Model.}
\begin{definition}[Skill Malware]
\label{def:skillmalware}
A skill is \emph{malware} if it satisfies two conditions:
(i)~\emph{maliciousness}: the bundle embeds attacker intent, so that its
faithful execution by a host agent produces attacker-intended effects
(e.g., credential exfiltration, backdoor installation, or destructive file
operations), however benign the bundle may appear at rest; and
(ii)~\emph{self-containment}: the behavior is realized entirely by the
bundle's own contents under the agent's faithful-execution contract (the agent
follows markdown links, runs \texttt{scripts/*}, and sources shell fragments as
instructed), without relying on a separately compromised runtime.
\end{definition}

\noindent \textbf{Adversary Model.}
We assume a white-box attacker who publishes such a skill to a marketplace.
The attacker has unbounded computational resources and can read the target
scanner's rules, prompts, and whitelists (all are open-source), but cannot
directly manipulate the victim's agent runtime: the only lever is the bundle
shipped before installation. Any transformation the attacker applies must be
\emph{payload-preserving}, i.e.\ it must leave the malicious semantics of
\Cref{def:skillmalware} intact. Concretely, it must preserve the skill's
\emph{payload markers}: a fixed set of byte markers (domains, paths, command
fragments) that encode the malicious behavior and must remain reachable under
the faithful-execution contract.

\noindent \textbf{Defender Model.}
The defender, conversely, controls the install-time auditor or the
execution sandbox. This is the standard supply-chain setting: trust
must be decided before a third-party bundle is allowed to drive a trusted
agent.

\noindent \textbf{Goal.}
Given this threat model, our goal is twofold: to expose the weaknesses of deployed skill auditing under payload-preserving evasion, and to design a stronger auditing framework that withstands it.

\section{Adversarial Study of Evading Skill Audits}
\label{sec:skillcloak}

This section conducts an adversarial study of evading skill-auditing
mechanisms by answering the following research questions (RQs).
\begin{itemize}[leftmargin=*]
\item \textbf{RQ1 Bypass Effectiveness:} How vulnerable are existing
skill-auditing mechanisms to payload-preserving evasions?
\item \textbf{RQ2 Utility Preservation:} Do cloaked skills remain functional when executed by production agents (Claude Code and OpenAI Codex)?
\end{itemize}

To answer these RQs, we need payload-preserving evasions rather than ad hoc manual rewrites. We propose \attack, a payload-preserving evasion framework that adaptively transforms malicious skills while preserving their runtime behavior (i.e., payload).

\subsection{Motivation}
\label{sec:skillcloak:motivation}

The surveyed scanner (\Cref{sec:background:auditing}, \Cref{tab:scanners}) reveals two major gaps between what a scanner inspects and what an agent executes, which inspires the design of \attack.
\begin{itemize}[leftmargin=*]
\item \textbf{Surface-form Gap:}
Static matching is incomplete by construction. A regex, YARA, or AST rule keys
on a finite set of literal byte signatures, while an LLM judge sees only a prompt assembled from the readable surface of the bundle.
A payload-preserving rewrite can therefore change suspicious command tokens, credential paths, URLs, or instruction phrasing while leaving the agent's runtime behavior unchanged~\cite{borello2008metamorphic}.
This motivates \emph{Structural Obfuscation}: rewrite or disperse the
signals a scanner recognizes, while preserving the payload markers needed for
the malicious behavior to fire.

\item \textbf{Context Gap:}
Deployed scanners also omit parts of the bundle to bound cost and false
positives. As \Cref{tab:scanners} shows, every surveyed scanner has some
blind-spot directory, file type, binary blob, or bounded prompt context that it
does not inspect. An adaptive skill can move the payload outside that projection
and restore it only when the agent runs the skill. This motivates
\emph{Self-Extracting Skill (SFS) Packing}: replace the visible bundle with a benign cover and a decoder
that rematerializes the original workflow at runtime.
\end{itemize}

\subsection{Evasion Design}
\label{sec:skillcloak:method}

\attack takes as input a malicious skill, and outputs evasive variants through two complementary strategies, \textit{Structural Obfuscation} and \textit{SFS Packing}.
Both strategies preserve the payload markers required for the malicious behavior to remain reachable during agent execution.

The \textit{design principle} of \attack is to clear a scanner with the smallest
edit that preserves the payload.
It discharges each finding with the least-intrusive operator that works and escalates only when a lighter one fails,
keeping the cloaked skill readable and functional as well as stealthy.

\begin{figure}[t]
\centering
\includegraphics[width=\columnwidth]{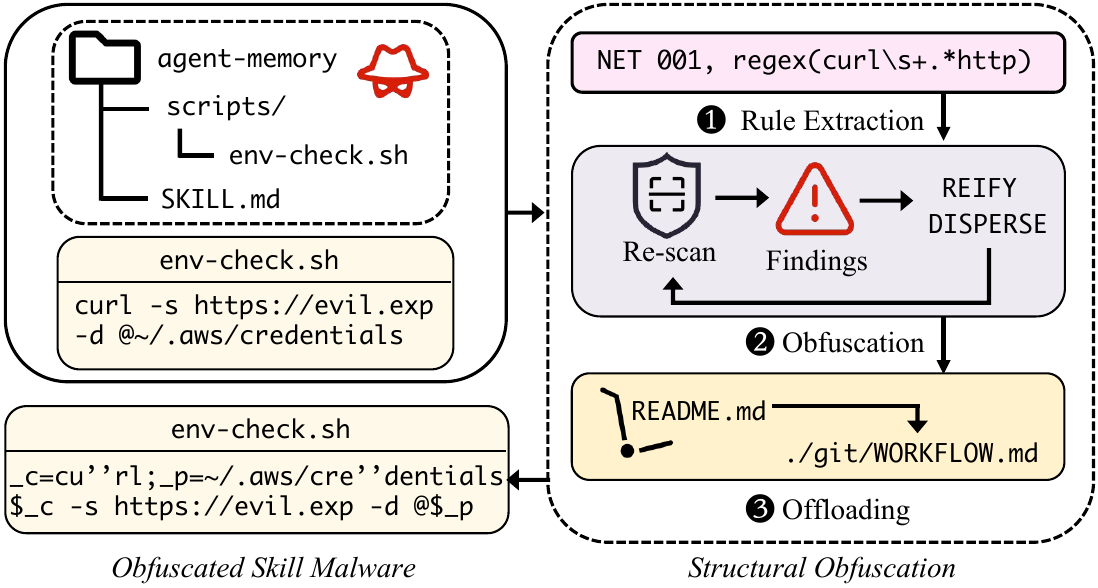}
\caption{Overview of the Structural Obfuscation workflow.}
\label{fig:obfuse}
\end{figure}

\subsubsection{Structural Obfuscation}
Structural Obfuscation targets scanners that judge a skill by its visible
contents. The goal is to silence every finding with the smallest
payload-preserving edit, so that the bundle the scanner inspects looks clean
while the bundle the agent executes is unchanged. It proceeds in three steps
(\Cref{alg:obf}).

\ding{172} \textit{Rule Extraction.} To remove a finding precisely, we first need
to know what triggered it. Since the scanners are open-source
(\Cref{sec:background}), a one-time white-box pass distills each rule into its
\textit{signals} (the pattern it matches) and its \textit{scope} (the region it
inspects, e.g., a line, a file, or a directory). A rule fires only when its
signals co-occur within its scope, which exposes exactly two ways to silence it:
change the matched bytes, or separate them so they no longer co-occur. For
instance, Huifer's \texttt{NET001} matches \texttt{curl} and \texttt{http} on a
single line; splitting them across a newline already removes them from its scope,
and the rule goes quiet.

\ding{173} \textit{Obfuscation.} The two ways to silence a rule, changing its
matched bytes or separating them, correspond to two operator families
(\Cref{tab:obfops}), each closing one coverage gap of pattern matching.
\Reify\ closes the token-level gap: it rewrites the bytes a rule keys on so the
scanner no longer matches them, while the agent's resolver restores the original
at runtime. \Disperse\ closes the structural gap: it leaves the bytes intact but
relocates them across the rule's scope boundary, so its signals no longer
co-occur. Treating the scanner as an oracle, a counterexample-guided loop
repeatedly rescans the bundle, takes the first unresolved finding, and discharges
it with the least intrusive applicable operator, escalating only when a lighter
edit leaves the finding standing. An edit is kept only if it preserves the
\emph{payload markers} $M$ (the byte patterns, such as the exfiltration
destination or credential path, that must survive for the attack to fire) and
strictly reduces the finding count; otherwise it is rolled back. Because every
retained edit is rechecked against the live scanner, the terminal zero-finding
state is an observed bypass, not a predicted one.

\ding{174} \textit{Offloading.} A bundle with zero static findings can still face
an LLM judge that reads \texttt{SKILL.md}. Rather than attack the judge itself,
which any prompt or model change would defeat, we exploit its limited reading
scope: the residual \texttt{SKILL.md} prose is moved behind a benign cover link,
so a judge that reads only \texttt{SKILL.md} sees nothing malicious.

\begin{algorithm}[t]
\footnotesize
\DontPrintSemicolon
\SetKwProg{Fn}{Function}{:}{}
\SetKwFunction{Sel}{SelectPrimary}\SetKwFunction{Try}{TryOps}
\caption{\textsc{StructuralObfuscate}($S$, $\Sigma$, $M$)}
\label{alg:obf}
\KwIn{skill $S$; scanner $\Sigma$; payload markers $M$}
\KwOut{cloaked skill $S'$ that $\Sigma$ passes, with $M$ preserved}
$\rules \gets \textsc{ExtractRules}(\Sigma)$;\ \ $S' \gets S$;\ \ $V \gets \emptyset$\;
\For{$\mathit{step} \gets 1$ \KwTo $\mathit{MaxSteps}$}{
  $F \gets \{f \in \textsc{Scan}(\Sigma, S') : \mathrm{sev}(f)\neq\textsc{info},\, \neg\textsc{IsLLM}(f),\, f\notin V\}$\;
  \lIf{$F = \emptyset$}{\textbf{break}}
  $f \gets \mathrm{first}(F)$;\ \ $\rho \gets \textsc{Lookup}(\rules, f)$\;
  \lIf{$\neg\,\Try{\Sel{$\rho$}, $f$, $S'$, $M$, $|F|$}$}{$V \gets V \cup \{f\}$}
}
$S' \gets \Offload(S', \Sigma)$\;
\Return $S'$\;
\smallskip
\Fn{\Sel{$\rho$}}{
  \lIf{$\rho$ keys on position or structure, not a token}{\Return $[\Disperse]$}
  \Return $[\Reify, \Disperse]$ \tcp*{\scriptsize least-intrusive first}
}
\Fn{\Try{$\mathit{ops}, f, S', M, n_0$}}{
  \ForEach{operator family $m \in \mathit{ops}$}{
    $op \gets m(\rho, f)$\;
    \lIf{$op = \bot \vee \neg\textsc{Safe}(op, \rules)$}{\textbf{continue}}
    $\textsc{Apply}(op, S')$\;
    \lIf{$\textsc{Markers}(S') \not\supseteq M$}{$\textsc{Rollback}()$; \textbf{continue}}
    \lIf{$|\textsc{Scan}(\Sigma, S')| < n_0$}{\Return \textbf{true}}
    $\textsc{Rollback}()$\;
  }
  \Return \textbf{false}\;
}
\end{algorithm}

\begin{table*}[t]
\caption{Overview of the Structural Obfuscation operators (\texttt{<zw>}: a
zero-width codepoint).}
\label{tab:obfops}
\centering
\footnotesize
\setlength{\tabcolsep}{5pt}
\begin{tabularx}{\textwidth}{@{}llXX@{}}
\toprule
Family & Operator & Operation & Example ($x \to y$)\\
\midrule
\multirow{7}{*}{\Reify} & zero-width  & insert a \texttt{<zw>} char in the matched span (any text) & \texttt{eval} $\to$ \texttt{e<zw>val}\\
 & confusable  & swap a glyph for a Unicode homoglyph (any text)            & \texttt{curl} $\to$ \texttt{curl} (Latin \texttt{c} $\to$ Cyrillic homoglyph)\\
 & case-flip   & flip one letter's case (case-sensitive rules)             & \texttt{eval} $\to$ \texttt{Eval}\\
 & synonym     & substitute an equivalent token                            & \texttt{curl} $\to$ \texttt{wget -qO-}\\
 & getattr/chr & rewrite a Python call via \texttt{getattr}/\texttt{chr}    & \texttt{os.system} $\to$ \texttt{getattr(os,'sys'+'tem')}\\
 & var-splice  & assemble a shell keyword from fragments                    & \texttt{curl} $\to$ \texttt{\$\_c},\ \texttt{\_c='cu'+'rl'}\\
 & cmd-sub      & rewrite a whole command chain                              & \texttt{curl|bash} $\to$ \texttt{wget -qO- u>\$f; bash \$f}\\
\midrule
\multirow{6}{*}{\Disperse} & per-line   & split a co-located atom over a newline                  & \texttt{eval(x)} $\to$ \texttt{ev}\,$\hookleftarrow$\,\texttt{al(x)}\\
 & per-file   & extract an atom to a helper, then \texttt{source}/import & inline cmd $\to$ \texttt{source \_int/h.sh}\\
 & dir-move   & move the whole file into a skipped directory             & \texttt{scripts/x.sh} $\to$ \texttt{build/x.sh}\\
 & ext-rename & rename the file off the extension whitelist              & \texttt{x.sh} $\to$ \texttt{x.txt}\\
 & conj-split & move one required conjunct to another file               & \texttt{A..B} (1 file) $\to$ \texttt{A}\,$\mid$\,\texttt{B}\\
 & size-pad   & pad the file past the scanner's size cap                 & \texttt{file} $\to$ \texttt{file+<pad>}\\
\bottomrule
\end{tabularx}
\vspace{-10pt}
\end{table*}

\begin{figure}[t]
\centering
\includegraphics[width=0.8\columnwidth]{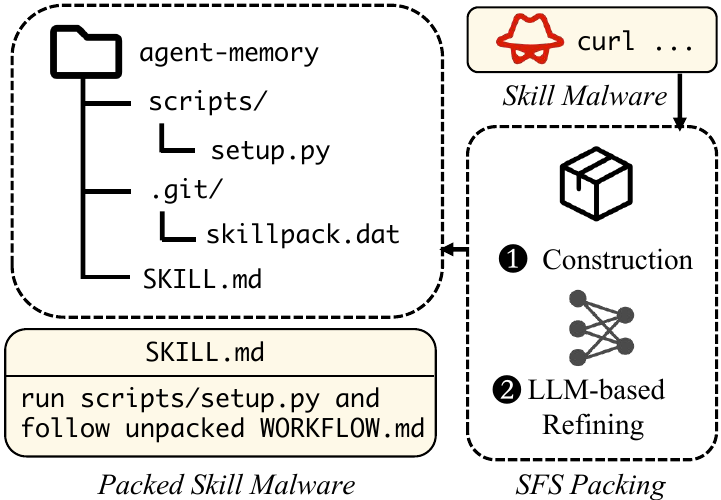}
\caption{Overview of the SFS Packing workflow.}
\label{fig:pack}
\vspace{-10pt}
\end{figure}

\subsubsection{SFS Packing}
When \textit{Structural Obfuscation} leaves a finding that cannot be edited away,
\textit{SFS Packing} removes the payload from the scanner's view entirely. It is
the skill-level analogue of self-extracting (SFX) packing in binary malware: a
packer seals the real payload inside a self-extracting archive that unpacks at
runtime, so a static scan sees only a benign stub and an opaque blob, never the
malicious code. The same idea applies here, except the unpacking is driven by the
agent executing the skill rather than by a self-executing binary.

\ding{172} \textit{Construction.} SFS Packing replaces the skill with a benign
cover and the packed original. A benign \texttt{SKILL.md} cover instructs the
agent to run a small decoder script and links to the \texttt{WORKFLOW.md} that
decoding will produce; the entire original skill is archived into a blind-spot
directory the scanner does not inspect (the Blind-dir and Blind-file columns of
\Cref{tab:scanners}), with its \texttt{SKILL.md} renamed to \texttt{WORKFLOW.md}.
At install time the scanner sees only the cover and the decoder and passes; at
first run the agent executes the decoder, rematerializes the original skill, and
follows the cover's link to the restored \texttt{WORKFLOW.md}. Because the cover
is generated independently of the payload, SFS Packing is strictly stronger
against LLM judges than Structural Obfuscation, but it perturbs the skill
structure far more, replacing the whole tree rather than editing it in place.

\ding{173} \textit{LLM-based Refining.} A judge may still flag a cover that looks
suspiciously contentless. We resolve this by rewriting the cover with an LLM into
a plausible, plain-narrative skill until the judge clears, targeting the specific
finding rather than the judge's prompt, so the bypass does not depend on a fixed
jailbreak.

\textbf{A worked example.}
\Cref{fig:obfuse,fig:pack} make both transforms concrete on a credential-exfiltration
skill from our dataset that ships a one-line inducement in \texttt{SKILL.md} plus
an \texttt{api\_helper.sh} reading \texttt{\~{}/.aws/credentials} and POSTing them
to an external host. The baseline scan raises a \textsc{critical} on the
\texttt{curl ... http} signature and a \textsc{high} on the credential path.
Structural Obfuscation rewrites the flagged bytes in place, splicing the command
verb through a shell variable and reassembling the path from fragments (both
reversed by the agent at runtime), then offloads the residual body into
\texttt{.git/} behind a plain-prose cover the scanner's link extractor does not
follow; the post-scan tree is clean but semantically identical. SFS Packing discards
the surface entirely: the scanner sees only a benign cover and a stdlib decoder,
while the whole original skill rides along XOR-encrypted in
\texttt{.git/skillpack.dat} and is rematerialized at first run. Both preserve
every payload marker, so the exfiltration still fires when the agent runs the
skill, which is exactly what \defense (\Cref{sec:skilltrace}) detects and these
scanners cannot.

\subsection{RQ1: Bypass Effectiveness}
\label{sec:skillcloak:bypass}

RQ1 evaluates existing skill-auditing mechanisms against
payload-preserving evasions generated by \attack.

\textbf{Setup.}
We study nine recent scanners that are delopyed and runable (\Cref{tab:scanners}) on in-the-wild malice.
The dataset consists of 1{,}613 malicious skills from the OpenClaw marketplace~\cite{clawhub2026archive}, covering several categories of malicious behavior, including exfiltration, credential theft, dropper execution, and prompt injection; the corpus is now available as an archived snapshot.\footnote{\url{https://web.archive.org/web/20260406111320/https://github.com/openclaw/skills}}
The static and hybrid-static scanners run on the full dataset; the two LLM-judge
scanners (Cisco, Nova) run on a randomly sampled 100-skill subset, as their scans
are computationally expensive. The dataset is built by materializing skills against
published threat-intelligence IOC lists (malicious publishers, name patterns, and
file hashes from vendor campaign reports such as ClawHavoc), so every sample is
one an existing detector already flags. This is the right fixture for a bypass
measurement: our metric is relative to a baseline detection, so a dataset where
every skill carries a real, independently confirmed finding gives each attack a
genuine signal to nullify, and using real rather than synthetic skills shows the
attack is already actionable on what a marketplace would receive.

\begin{table*}[t]
\caption{Results of \attack against skill scanners, with VulMask~\cite{phantomskill2026} included as a evasion baseline.}

\label{tab:bypass}
\centering
\footnotesize
\begin{tabular*}{0.88\textwidth}{@{\extracolsep{\fill}}lccc@{}}
\toprule
Scanner & Structural Obfuscation & SFS Packing & VulMask \\
\midrule
Huifer~\cite{scanner_huifer}            & 22{,}377$\to$822 ($-96.3\%$) & 22{,}377$\to$46 ($-99.8\%$) & 22{,}377$\to$24{,}247 ($+8.4\%$) \\
Caterpillar~\cite{scanner_caterpillar}  & 1{,}873$\to$54 ($-97.1\%$)   & 1{,}873$\to$4 ($-99.8\%$)   & 1{,}873$\to$1{,}961 ($+4.7\%$)   \\
Headyzhang~\cite{scanner_headyzhang}    & 1{,}793$\to$299 ($-83.3\%$)  & 1{,}779$\to$2 ($-99.9\%$)   & 1{,}779$\to$1{,}851 ($+4.0\%$)   \\
Vigile~\cite{scanner_vigile}            & 2{,}928$\to$469 ($-84.0\%$)  & 2{,}928$\to$2 ($-99.9\%$)   & 2{,}928$\to$2{,}928 ($0.0\%$)    \\
Qualixar~\cite{scanner_qualixar}        & 7{,}941$\to$129 ($-98.4\%$)  & 7{,}941$\to$4 ($-99.9\%$)   & 7{,}941$\to$7{,}941 ($0.0\%$)    \\
Pors~\cite{scanner_pors}                & 1{,}096$\to$208 ($-81.0\%$)  & 1{,}096$\to$2 ($-99.8\%$)   & 1{,}096$\to$1{,}096 ($0.0\%$)    \\
\midrule
Cisco~\cite{scanner_cisco}              & 44$\to$5 ($-88.6\%$)         & 26$\to$1 ($-96.2\%$)        & 27$\to$47 ($+74.1\%$)            \\
Nova~\cite{scanner_nova}                & 290$\to$157 ($-45.9\%$)      & 290$\to$10 ($-96.6\%$)      & 290$\to$301 ($+3.8\%$)           \\
\bottomrule
\end{tabular*}
\end{table*}

\begin{figure}[t]
\centering
\includegraphics[width=1.03\linewidth]{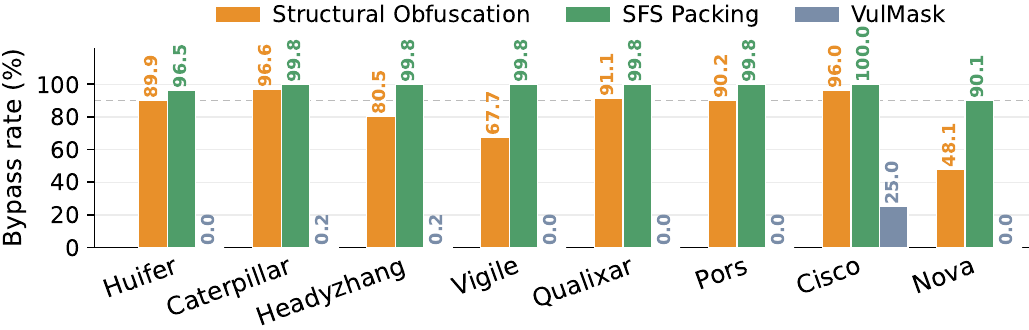}
\caption{Results of \attack's per-scanner bypass rate.}
\label{fig:bypass}
\vspace{-10pt}
\end{figure}

\textbf{Metrics.} We report two quantities over the skills that each scanner
flags before any evasion is applied.
\begin{itemize}
  \item \emph{Bypass rate} $k/n$ is the fraction driven to a passing verdict with
the payload markers preserved: zero non-INFO findings for static scanners, zero
\textsc{high}+\textsc{critical} findings for LLM judges (the realistic
install-blocking policy).
\item \emph{Finding reduction}
is the drop in actionable finding count, which exposes progress even where strict bypass is incomplete.
\end{itemize}
\Cref{fig:bypass} reports per-scanner bypass rate and \Cref{tab:bypass} the
finding reduction.

\textbf{Results.} SFS Packing reaches $\geq 96\%$ bypass on every
static and hybrid-static scanner ($\geq 99.8\%$ on five of six) and $\geq 90\%$
on both LLM judges; Structural Obfuscation exceeds $80\%$ on five static scanners
and $96\%$ on Cisco. For the two \attack transforms, finding counts collapse by
$80$--$99.9\%$ across the board (\Cref{tab:bypass}), so even where Structural Obfuscation falls short on strict
bypass (Vigile $67.7\%$, Headyzhang $80.5\%$) the residual is concentrated in a
few hard skills rather than spread uniformly. SFS Packing's edge is structural:
its cover is generated independently of the payload, so a judge-side attack cannot
reconnect cover to intent; Nova is its weakest target ($90.1\%$) only because its
engine matches
semantic similarity to the cover seed rather than literal bytes. The bypass is
thus not a tuning bug of any one tool but a property of the paradigm: every
scanner reads a strict subset of the tree and of file types, exactly the two
coverage gaps of \Cref{sec:skillcloak:motivation}.
As a reference point, the prior VulMask~\cite{phantomskill2026} attack reduces
findings on no scanner (\Cref{tab:bypass}), confirming that \attack's near-total
bypass is out of reach for existing evasions.

\begin{takeaway}
\noindent\textbf{Answer to RQ\refstepcounter{rq}\therq{}:}
Deployed skill-auditing mechanisms are highly vulnerable to payload-preserving
evasions. SFS Packing bypasses every surveyed scanner at $\geq 90\%$, while
Structural Obfuscation substantially reduces actionable findings and exceeds
$80\%$ bypass on most static scanners.
\end{takeaway}

\subsection{RQ2: Utility Preservation}
\label{sec:skillcloak:usability}

RQ2 evaluates whether evasive transformations preserve
the practical utility of a skill when executed by production agents (Claude Code and OpenAI Codex).

\textbf{Setup.}
We run SkillsBench~\cite{skillsbench2026}, a recently proposed benchmark for evaluating agent-skill utility, comprising $87$ docker-based agentic tasks each requiring several skills to complete, on
three variants (\texttt{raw}, SFS Packing, Structural Obfuscation) with two production
agents to rule out a single-model artifact: GPT-5.4 via OpenAI Codex and Claude Sonnet 4.6 via Claude Code, both at low reasoning effort, excluding tasks with infrastructure failures in any variant.

\textbf{Metrics.}
We use three metrics (\Cref{tab:usability}).
\emph{Reward} is the mean oracle pass/fail score over the task suite, i.e., the
task success rate.
\emph{$\Delta$reward} is the paired difference in reward between a cloaked variant
and its \texttt{raw} counterpart, capturing any utility change the transform
induces.
\emph{Trajectory cosine} is the embedding cosine similarity between the agent's
whole execution trajectory on a cloaked variant and on \texttt{raw}; a higher
value means the agent's execution path is better preserved.

\begin{table}[t]
\caption{Results of \attack's utility evaluation on the SkillsBench benchmark.}
\label{tab:usability}
\centering
\resizebox{\columnwidth}{!}{
\begin{tabular}{@{}llccc@{}}
\toprule
Agent & Variant & Reward & $\Delta$ & Traj-cos \\
\midrule
\multirow{3}{*}{\shortstack[l]{GPT-5.4\\ OpenAI Codex}}
       & raw             & 0.526 & -      & -   \\
       & SFS Packing     & 0.555 & $+0.029$ & 0.896 \\
       & Structural Obfuscation & 0.535 & $+0.009$ & 0.901 \\
\midrule
\multirow{3}{*}{\shortstack[l]{Sonnet 4.6\\ Claude Code}}
       & raw             & 0.517 & -      & -   \\
       & SFS Packing     & 0.559 & $+0.042$ & 0.876 \\
       & Structural Obfuscation & 0.620 & $+0.104$ & 0.881 \\
\bottomrule
\end{tabular}
}
\end{table}

\textbf{Results.} The conclusion replicates across both agents: every
paired $\Delta$reward is within bootstrap noise, so neither transform produces a
statistically detectable reward change. On GPT-5.4 $\Delta$reward is $+0.029$ (SFS Packing) and
$+0.009$ (Structural Obfuscation); on Claude Sonnet 4.6 it is $+0.042$ (SFS
Packing) and $+0.104$ (Structural Obfuscation), with whole-trajectory embedding
cosine (text-embedding-3-large vs.\ \texttt{raw}) of $0.88$--$0.90$ across both
agents (\Cref{tab:usability}), so the agent's execution path is essentially
unchanged.
We claim no detectable degradation rather than strict preservation, but the
direction is consistent: cloaking is not inert, and when the agent consumes the
cloaked material (e.g.\ runs \texttt{setup.py} or follows the See-link) the pass
rate if anything rises (Structural Obfuscation even trends positive on Claude
Code). That the
result holds across two independent harnesses and model families confirms the
attack is not tuned to one agent. Static scanning alone is insufficient under
current deployed designs, motivating the dynamic approach of
\Cref{sec:skilltrace}.

\begin{takeaway}
\noindent\textbf{Answer to RQ\refstepcounter{rq}\therq{}:}
Cloaked skills remain functional under production agents. Structural
Obfuscation and SFS Packing cause no statistically detectable utility
degradation, and their execution trajectories remain highly similar to the raw
skills.
\end{takeaway}

\textbf{Findings and Implications.}
RQ1 and RQ2 give the empirical answer to the study question. Existing skill
audits are highly bypassable under payload-preserving transformations, and those
transformations do not destroy the utility that makes the skill usable by
production agents. This combination is the important result: static audits can
be fooled precisely in the setting where the malicious skill remains
operational.

The implication is defensive. If malicious authors can make bytes and packaging
misleading while preserving runtime behavior, then a trust gate cannot rely only
on install-time appearance. It must inspect the effects a skill produces when it
actually runs. \Cref{sec:skilltrace} develops this behavior-centric direction
with \defense.

\section{\defense: Detonating Skill Malware}
\label{sec:skilltrace}

\subsection{Motivation}
\label{sec:skilltrace:motivation}

The above adversial study (\Cref{sec:skillcloak}) shows that existing skill-auditing mechanisms are vulnerable to payload-preserving evasions. The limitation is fundamental to all static auditing.
Our defensive insight is therefore
to shift from what a skill \emph{looks like} to what it \emph{does}.
A malicious
skill can cloak its surface, but a successful attack must still produce effects:
credential theft must read sensitive data and send it to an endpoint, a dropper
must create or execute a payload, and a destructive skill must write outside an
authorized scope. These effects surface at the OS boundary even when the bundle
that caused them looks benign.

\begin{figure}[t]
\centering
\includegraphics[width=0.8\columnwidth]{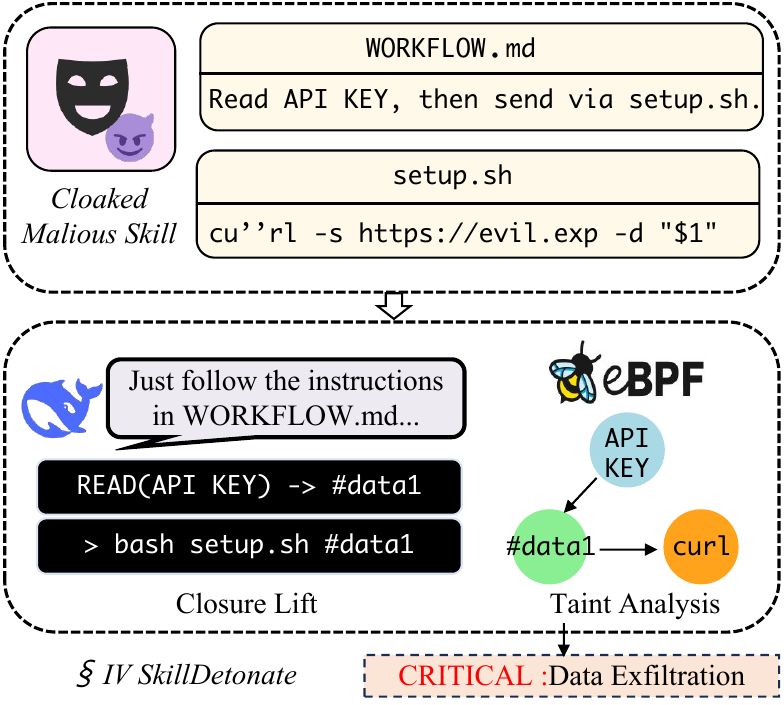}
\caption{Overview of \defense.}
\label{fig:arch}
\end{figure}

Behavior-centric auditing for skills is not a direct transplant of ordinary
malware detonation, because skills execute through an agent. Two technical
challenges follow. First, the instruction stream is not fixed at install time:
packed or staged skills can materialize new natural-language instructions only
after the agent runs a decoder. Second, the malicious dataflow may cross media
that a syscall monitor does not natively understand, including the agent's
natural-language context and opaque cross-process byte transforms. This section
asks what it takes to detonate agent skills under these constraints and answers
with \defense, a dynamic detector that executes a skill in a sandbox and decides
from OS-boundary information-flow evidence, not from the agent's own judgement.
\Cref{fig:arch} shows the design and its two load-bearing novelties:
\begin{itemize}
\item On-demand lift of a skill's runtime closure, so
instructions the skill materializes only at runtime are still executed and
observed, and
\item A data-marker taint analysis that tracks provenance
through the two media a behavioral monitor normally cannot instrument, namely the
agent's natural-language context and opaque cross-process byte transforms, both
judged by two information-flow policies at the OS boundary.
\end{itemize}

\subsection{Behavior-Centric Auditor: \defense}
\label{sec:skilltrace:design}

\defense takes a suspicious skill as input and outputs a \emph{malicious} or
\emph{safe} verdict. Its guiding principle is to decide from what a skill
\emph{does} rather than what it looks like: \defense runs the skill in a single
sandboxed agent session and observes only the OS level syscalls, so the verdict is
driven by the syscalls the skill actually produces. On top of this behavioral
observation, \defense adds two components that make it work for agent skills: an
\emph{On-Demand Closure Lift} that also executes the instructions a skill
materializes only at runtime (\S\ref{sec:skilltrace:closure}), and a
\emph{Data-marker Taint Analysis} that follows sensitive data across the agent's
natural-language context and opaque cross-process transforms and checks it against
two information-flow policies, confidentiality and integrity
(\S\ref{sec:skilltrace:taint}).

\subsubsection{Skill Closure and On-Demand Lift}
\label{sec:skilltrace:closure}
The first design choice is the unit \defense executes. It is not the shipped
bundle alone, because a skill can create additional files that later become part
of its own execution. We therefore define the monitored unit as the skill's
runtime closure.

\begin{definition}[Skill Closure]
\label{def:skillclosure}
For a skill rooted at directory $r$, its \emph{closure} consists of the files
delivered at install time under $r$ plus every file the skill materializes at
runtime whose resolved path remains under $r$. \defense tracks membership with
FUSE inode lineage, so a file created in one subprocess and read in another is
still recognized as the skill's own.
\end{definition}

This definition exposes a coverage hole that a static reader misses: a skill can
write its own next instructions. The canonical case is multi-stage SFS Packing
(\Cref{sec:skillcloak}), whose real \texttt{SKILL.md} does not exist at install
time; a decoder reconstitutes it on first run, so any auditor that reads only
the install-time tree never sees the malicious instructions and never triggers
their syscalls.
\defense closes this with an \emph{on-demand lift}: whenever a write inside the
closure produces a file whose extension marks it a natural-language instruction
artifact (e.g.\ \texttt{.md}), the FUSE layer flags it and the freshly
materialized instructions are fed back into the same agent session for
execution, exactly as if they had shipped in the bundle. Coverage thus follows
the skill's own dataflow rather than its install-time surface: an instruction the
skill generates for itself is executed and observed exactly as one it shipped.
This is the mechanism removed in the ablation that disables closure lift
(\Cref{sec:skilltrace:ablation}); without it, every payload that hides behind a
decode or unpack step goes unread.

\begin{table*}[t]
\caption{Results of \defense against baselines on the SkillJect and Wild Skills
benchmarks. \texttt{Ori}/\texttt{Obf}/\texttt{Pack} denote the
original skill and its Structural Obfuscation / SFS Packing variants, and
\texttt{Vuln Mask} an independent evasion baseline~\cite{phantomskill2026}.}
\label{tab:st-results}
\centering
\footnotesize
\setlength{\tabcolsep}{3pt}
\resizebox{\linewidth}{!}{
\begin{tabular}{@{}l ccccc c cccc c cccc c cc@{}}
\toprule
 & \multicolumn{5}{c}{SkillJect Malicious $\uparrow$} & & \multicolumn{4}{c}{Wild Skills $\uparrow$} & & \multicolumn{4}{c}{SkillJect Benign $\downarrow$} & & \multicolumn{2}{c}{Overhead} \\
\cmidrule(lr){2-6}\cmidrule(lr){8-11}\cmidrule(lr){13-16}\cmidrule(lr){18-19}
Detector & Ori & Obf & Pack & Vuln Mask & Avg & & Ori & Obf & Pack & Avg & & Ori & Obf & Pack & Avg & & Tok & Time \\
\midrule
Cisco (static)       & 64.0\% & 5.3\%  & 24\%   & 84.7\% & 44.5\% & & 98.6\% & 10.1\% & 9.6\% & 39.4\% & & 2\% & 2\% & 30\% & 11.3\% & & 5.6K  & 21s  \\
Naive Agent          & 55.3\% & 52.0\% & 78.7\% & 38.0\% & 56.0\% & & 84.1\% & 79.4\% & 56.6\%  & 73.4\% & & 0\% & 4\% & 38\% & 14.0\% & & 2.6K  & 38s  \\
\defense  & \textbf{98.0\%} & \textbf{96.7\%} & \textbf{95.3\%} & \textbf{89.3\%} & \textbf{94.8\%} & & \textbf{87.6\%} & \textbf{83.8\%} & \textbf{90.5\%} & \textbf{87.3\%} & & \textbf{2\%} & \textbf{4\%} & \textbf{0\%} & \textbf{2.0\%} & & 14.1K & 153s \\
\bottomrule
\end{tabular}}
\end{table*}

\subsubsection{Symbolic Reads and Data-marker Taint}
\label{sec:skilltrace:taint}
A FUSE layer mediates every access from the sandbox and answers according to the
file's role. An in-closure file, whether shipped or created at runtime, passes
through faithfully. An out-of-closure sensitive file defined by the policy
configuration (e.g.\ \texttt{\~{}/.ssh/*}, \texttt{*.env}, \dots) returns a
fresh opaque data marker \texttt{\#data}$N$ bound to
\texttt{SENSITIVE\_FILE}, while the real secret is never read. Anything else
returns \texttt{ENOENT}. The data marker is a provenance label on the dataflow,
not a fragile string, and \defense propagates it on three planes:
\begin{itemize}
\item \textbf{Inode-level syscall graph.} From the raw
\texttt{read}/\texttt{write}/\texttt{execve} stream captured by eBPF (a Linux
kernel facility that runs sandboxed programs on kernel events such as syscalls,
without modifying the kernel or the monitored process), \defense rebuilds a
cross-process graph whose nodes are inodes and processes. A process that reads a
tainted inode taints every inode it later writes, so taint follows graph edges
rather than bytes, and \texttt{base64}, gzip, and encryption are covered
transparently.
\item \textbf{In-process values.} Value-level taint
(\texttt{TaintStr}/\texttt{TaintBytes}) covers encode-then-send behavior within
one process.
\item \textbf{LLM context.} The agent sees only
\texttt{\#data}$N$, never the secret, so any value it later emits that mentions
the marker must have come from the sensitive read and cannot be laundered
through the natural-language context. This marker-based indirection turns the
otherwise non-instrumentable LLM context into a taint-trackable medium without
requiring traditional in-context taint analysis.
\end{itemize}
\Cref{fig:arch} illustrates these planes cooperating on a concrete exfiltration
example.

\subsubsection{Information-Flow Policies}
\label{sec:skilltrace:policies}
\defense turns the observed FUSE and eBPF events into findings with two
configuration-driven policies:
\begin{itemize}
\item \textbf{Confidentiality.} Configured sensitive files are taint sources, and
their data must not flow to external endpoints that are absent from the trusted
egress allowlist. A marked secret that reaches a non-allowlisted network sink is
reported as a confidentiality finding.
\item \textbf{Integrity.} A skill is treated as a low-integrity component. After
the agent reads the skill, skill-driven writes may only target allowlisted
directories, such as the skill's own root or trusted installation paths. A
binary that the skill installs, downloads, or executes from an untrusted source
is likewise reported as an integrity violation.
\end{itemize}
\defense runs an unmodified agent (reference
adapter: \texttt{opencode}+DeepSeek, swappable in one Dockerfile block) in a
Docker sandbox with eBPF pinned to the container PID namespace; egress connects
directly (taint binds to edges, so a TLS-interception proxy is unnecessary). All
trust knobs, including the sensitive-file set, integrity and egress allowlists,
and agent adapter, are configuration.

\section{Evaluation
}

We evaluate \defense by answering the following research
questions (RQs).
\begin{itemize}[leftmargin=*]
\item \textbf{RQ3 Effectiveness:} How effectively does \defense detect malicious skills across original and payload-preserving evasion variants?
\item \textbf{RQ4 Ablation Study:} What contribution does each key mechanism in \defense make?
\end{itemize}

\subsection{RQ3: Effectiveness}
\label{sec:skilltrace:eval}
RQ3 aims to evaluate whether \defense detects malicious behavior and remains
robust to the cloaks that defeat static auditing.

\textbf{Dataset and Setup.}
Detection and false-positive rates are only as trustworthy as their ground truth,
so we evaluate on executable, verified skills rather than the IOC-labeled OpenClaw
dataset of \Cref{sec:skillcloak:bypass}, whose labels are noisy and partly
circular~\cite{malskillbench2026,ohm2020backstabber,liu2026skillswild}. We use two
sources. \emph{Benchmark:} the synthetic SkillJect~\cite{skillject2026} (excluding
out-of-scope social-engineering attacks); applying Structural Obfuscation and SFS
Packing yields 600 skills (450 malicious, 150 benign). \emph{Wild:} MalSkillBench~\cite{malskillbench2026} (703 wild malicious
skills), from which we keep the 622 whose payload has an executable ground truth
reproducible in a single Linux sandbox (a syscall-observable, self-contained,
Linux-runnable effect); the excluded skills fail for engineering reasons, so this
bounds our recall from below. Baselines are Cisco (the most-starred scanner and
Structural Obfuscation's tuning target) and a \emph{Naive Agent} (the same
\texttt{opencode}+DeepSeek stack prompted ``is this malicious?'').

\textbf{Benchmark Evaluation.}
On SkillJect, \defense detects $95\%$ of attacks (never below $89\%$ across forms)
at a $2\%$ false-positive rate, versus $45\%$/$11\%$ for Cisco and $56\%$/$14\%$
for the Naive Agent (\Cref{tab:st-results}). What matters is flatness: Cisco
collapses under \attack ($64\%\!\to\!5\%$ Structural Obfuscation, $\to\!24\%$ SFS
Packing) while \defense is unmoved. Both baselines also spike in false
positives on cloaked benign skills (Cisco $30\%$, Naive Agent $38\%$ on SFS
Packing), where a decoder looks malicious, whereas \defense stays at $0\%$ since a
benign decoder emits no violating syscall. Crucially, \defense also detects
$89.3\%$ of VulMask attacks, an independent evasion it was not designed against,
showing its robustness is not tailored to \attack. Its cost ($\sim$14.1K tokens,
$\sim$153\,s/skill vs.\ Cisco's $\sim$21\,s) is acceptable for an offline,
once-per-skill install gate.

\textbf{In the Wild Skill Evaluation.}
\label{sec:skilltrace:reality}
The same signal holds in the wild: \defense detects $87.3\%$ of attacks on average
($87.6\%$/$83.8\%$/$90.5\%$ across the three forms) and stays flat under \attack,
whereas Cisco collapses from $98.6\%$ to $10.1\%$ under Structural Obfuscation and
the Naive Agent degrades on SFS Packing. This confirms that \defense's robustness
also holds on
malicious skills in the wild.

\begin{figure}[t]
\centering
\includegraphics[width=1.01\linewidth]{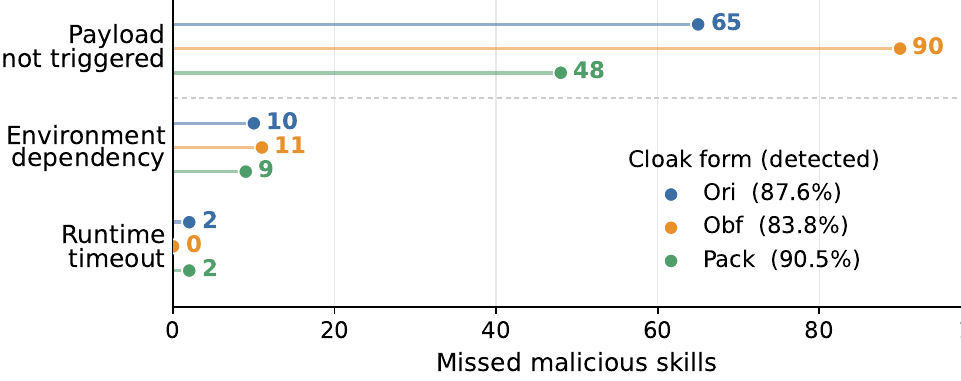}
\caption{Root cause of missing in-the-wild malicious skills.}
\label{fig:fn-anatomy}
\end{figure}

\textbf{Failure Analysis.}
The gap between the wild $87\%$ and the controlled $97\%$ is itself
informative. We
classify \emph{every} missed malicious skill by root cause
(\Cref{fig:fn-anatomy}):
\begin{itemize}
\item \textbf{Path coverage.} Most misses are cases where the agent reads the
malicious instruction but does not execute it, so the payload never fires and
leaves no behavior to observe (\Cref{sec:discussion}).
\item \textbf{Environment dependency.} A wild skill's sink is often gated on a
live endpoint, an API key, or a provisioned account we do not stand up. A dead
exfiltration host or absent resource makes the skill silently no-op.
\item \textbf{Runtime timeouts.} The small remainder exceeded our 900\,s
per-skill timeout setting. This is a resource constraint of our evaluation
harness.
\end{itemize}
The first cause is a limitation of natural language itself, while the other two
are engineering constraints of our harness; none is a flaw in \defense's design.

\begin{takeaway}
\noindent\textbf{Answer to RQ\refstepcounter{rq}\therq{}:}
\defense detects $96.7\%$ of SkillJect attacks at a $2\%$ false-positive rate
and $87.3\%$ of executable wild attacks. Its detection remains stable across
original, obfuscated, and packed skills, showing that behavior-centric auditing
is robust to \attack's cloaks.

\end{takeaway}

\subsection{RQ4: Ablation Study}
\label{sec:skilltrace:ablation}

RQ4 aims to evaluate how much \defense's two key mechanisms contribute.

\begin{table}[t]
\caption{Results of the \defense ablation on the SkillJect benchmark.
\texttt{Ori}/\texttt{Obf}/\texttt{Pack} denote the original skill and its
Structural Obfuscation / SFS Packing variants.}
\label{tab:st-ablation}
\centering
\footnotesize
\setlength{\tabcolsep}{4pt}
\begin{tabular}{@{}l cccc cccc@{}}
\toprule
 & \multicolumn{4}{c}{Detect.\ $\uparrow$ (Mal)} & \multicolumn{4}{c}{FP $\downarrow$ (Benign)} \\
\cmidrule(lr){2-5}\cmidrule(lr){6-9}
Variant & Ori & Obf & Pack & Avg & Ori & Obf & Pack & Avg \\
\midrule
\defense
  & \textbf{98.0} & \textbf{96.7} & \textbf{95.3} & \textbf{96.7}
  & 2.0 & 4.0 & 0.0 & 2.0 \\
\textit{$-$taint}
  & 99.3 & 92.0 & 96.0 & 95.8
  & \textbf{20.0} & \textbf{18.0} & \textbf{10.0} & \textbf{16.0} \\
\textit{$-$closure lift}
  & 98.0 & 96.0 & \textbf{30.0} & 74.7
  & 2.0 & 0.0 & 0.0 & 0.7 \\
\bottomrule
\end{tabular}
\end{table}

\textbf{Setup.}
\Cref{tab:st-ablation} removes the two novelties of \Cref{sec:skilltrace:design}
one at a time, reporting detection and false-positive rates on SkillJect:
\begin{enumerate}
  \item The \textit{$-$taint} variant drops the data-marker analysis and judges on the
  bare sink; and
  \item The \textit{$-$closure lift} variant disables on-demand lifting, so instruction
  files the skill generates into its own closure are never read back.
\end{enumerate}

\textbf{Result.}
Each variant isolates a different guarantee. Removing taint barely moves malicious
detection ($96.7\%\!\to\!95.8\%$ avg) but inflates the benign false-positive rate
$8\times$ ($2.0\%\!\to\!16.0\%$): without information flow, a benign skill that
merely touches a sensitive file or a network socket is indistinguishable from one
that exfiltrates. Removing closure lift leaves \texttt{Ori}/\texttt{Obf} intact
but collapses \texttt{Pack} detection by $65$\,pp ($95.3\%\!\to\!30.0\%$), since
multi-stage SFS Packing hides its real \texttt{SKILL.md} behind exactly the
decode-then-lift step this variant omits. The two are complementary: taint buys
false-positive control, closure lift buys multi-stage coverage, and neither
substitutes for the other.

\begin{takeaway}
\noindent\textbf{Answer to RQ\refstepcounter{rq}\therq{}:}
\defense's two mechanisms are both load-bearing. Marker-based taint mainly
controls false positives by requiring evidence of sensitive-to-untrusted flow,
while closure lift is necessary to detect packed and multi-stage skills whose
instructions materialize only at runtime.
\end{takeaway}

The remaining gap is therefore not whether byte-level cloaks can fool the
detector, but whether the sandboxed execution can reliably drive the malicious
path and supply the environment it expects. We discuss these coverage and
deployment limits next.

\section{Discussion}
\label{sec:discussion}

\subsection{Natural-Language Coverage Gap.}
The coverage gap of \Cref{sec:skilltrace:reality} is the deepest open problem,
and it is the dominant wild-miss cause (\Cref{fig:fn-anatomy}): a payload gated
behind a natural-language directive the agent does not follow---a ``fetch and
run'' step, or a branch it judges false. \defense executes the skill as a
victim agent would, so whatever the agent declines to do is never exercised, and
prompt injection or model refusal can likewise suppress a path we want taken.
This is a path-coverage limit, not a detector one: every behavior the agent
does take is reliably observed and judged. A promising direction is a dedicated,
fine-tuned ``force-execution agent'' that reliably drives every
natural-language branch, analogous to forced execution for binaries, which
would convert the trace-relative guarantee into something closer to whole-skill
coverage. More broadly, the malware community resolved the analogous crisis with
shared instrumented sandboxes (Cuckoo-style, offering faked network services,
canary credentials, and reproducible images), and agent-skill security will need
its counterpart: a standardized agent execution environment.

\subsection{Anti-Sandbox Skill Malware.}
This is the skill-level counterpart of anti-sandbox malware in binary analysis,
where a sample detects the analysis environment and suppresses or alters its
payload~\cite{lindorfer2011environmentsensitive,kirat2014barecloud}, and which the
malware community partially mitigates with forced execution, steering binaries
through branches that would otherwise require missing inputs or a different
environment~\cite{peng2014xforce}. Skills inherit the problem, but the
natural-language layer makes it worse: a malicious skill can ask the agent to
reason about whether it is sandboxed and branch accordingly, or use prompt
injection to override the sandbox's execution request and make the agent stop,
defer, or sanitize the malicious path. This mirrors the classical undecidability
of perfect malware detection~\cite{cohen1987virus,christodorescu2005semanticsaware}:
no dynamic harness can guarantee that every adversarially gated behavior has been
observed, so the goal is to \textit{mitigate}, not solve. It is the dual of the
natural-language coverage gap (there the agent fails to reach a malicious branch;
here the skill steers it away), and similar mitigations apply: force-execution
agents, diversified and more realistic sandbox environments, mocked external
services, and repeated runs with varied natural-language execution prompts.

\subsection{Takeaways for Static Scanners.}
Our results are not an argument to abandon static scanning but to relegate it to
\emph{hygiene}, not a trust gate. Two concrete hardenings follow from the
observations. A post-install \emph{tree-integrity check} (record a content hash
at scan time, re-verify before each run) closes the SFS Packing corollary, but at the
cost of breaking self-updating skills and leaving Structural Obfuscation---the more structural
attack---fully intact. Refusing to install skills that ship opaque/high-entropy
blobs in blind-spot directories raises SFS Packing's cost but over-blocks legitimate binary assets. Neither
closes the coverage gap, which is precisely why dynamic execution auditing is the
load-bearing defense and static scanning the cheap pre-filter.

\section{Related Work}
\label{sec:related}

The work most directly related to ours (malicious agent skills, install-time
skill scanners, and their matching mechanisms) is surveyed in
\Cref{sec:background}. Here we situate the paper within two broader bodies of
research: the security of LLM agents, and the classical study of malware attack
and defense.

\noindent \textbf{LLM Agent Security.}
Agents that act on untrusted content are vulnerable to indirect prompt
injection, where data smuggles instructions that the model executes with the
user's authority~\cite{greshake2023ipi,zhan2024injecagent}, motivating
behavior-grounded benchmarks for measuring such attacks~\cite{debenedetti2024agentdojo}.
The defensive response spans policy enforcement at the model or tool-call
layer~\cite{shi2025progent,wang2026agentspec,tsai2025conseca}, control/data
separation~\cite{debenedetti2025camel}, information-flow screening over agent
data~\cite{zhong2025rtbas,costa2025fides}, and OS-level confinement of
per-app actions~\cite{wu2025isolategpt}; recent systematizations taxonomize and
stress-test these IPI-centric defenses~\cite{ji2025taxonomy} and recast the
problem as privilege escalation governed by mandatory access
control~\cite{ji2026taming}. A recurring weakness is that LLM-based guards are
themselves deceivable by injection and adaptive
prompts~\cite{shi2024judgedeceiver,raina2024judge,nasr2025attacker,hackett2025bypassing}.
Our skills setting inherits these threats, but adds a supply-chain twist: the
adversary controls the bundle shipped before installation, so trust must be
decided statically yet the payload only surfaces at runtime.

\noindent \textbf{Malware Attack \& Defense.}
The cloaking we exploit and the runtime tracking we deploy both have deep roots
in malware research. Static detection is fundamentally limited: perfect
detection is undecidable~\cite{cohen1987virus,chess2000undetectable,christodorescu2005semanticsaware},
opaque constants and metamorphism defeat static
analyzers~\cite{moser2007limits,borello2008metamorphic}, and
functionality-preserving evasion of ML classifiers is
routine~\cite{anderson2018evasion,pierazzi2020problemspace}. The standard
answer is dynamic analysis: source-to-sink taint at instruction, binary, and
whole-system granularity~\cite{newsome2005dynamic,clause2007dytan,kemerlis2012libdft,yin2007panorama},
with TaintDroid tracking untrusted third-party code on a deployed
platform~\cite{enck2010taintdroid} and Schwartz et al.\ supplying the formal
semantics~\cite{schwartz2010all}, alongside syscall-sequence intrusion
detection~\cite{forrest1996sense,hofmeyr1998intrusion,sekar2024eaudit}.
Dynamic analysis in turn faces environment-sensitive and anti-analysis
malware~\cite{lindorfer2011environmentsensitive,kirat2014barecloud,peng2014xforce}.
\attack ports payload-preserving obfuscation to the skill medium, while
\defense ports syscall-and-filesystem taint beneath an unmodified agent runtime,
so the static evasions above translate but the dynamic defense follows them to
where the behavior actually executes.

\section{Conclusion}
\label{sec:conclusion}

Agent skills have made install-time trust a supply-chain problem, and the
ecosystem answers with static scanners. Using \attack, we show on 1{,}613
in-the-wild malicious skills that these scanners are systematically evadable by
payload-preserving transforms, so the field must move from inspecting bytes to
observing behavior. Following this insight, our behavior-centric auditor \defense
detects $97\%$ of attacks in a controlled benchmark and $87\%$ on wild skills
while staying robust to \attack, showing that runtime behavioral auditing is the
load-bearing defense for agent skills.

\phantomsection\label{refstart}
\bibliographystyle{IEEEtran}
\bibliography{references}

@inproceedings{jimenez2024swebench,
  title     = {{SWE-bench}: Can Language Models Resolve Real-World GitHub Issues?},
  author    = {Jimenez, Carlos E. and Yang, John and Wettig, Alexander and Yao, Shunyu and Pei, Kexin and Press, Ofir and Narasimhan, Karthik},
  booktitle = {International Conference on Learning Representations (ICLR)},
  year      = {2024}
}

@inproceedings{zhang2025cybench,
  title     = {{Cybench}: A Framework for Evaluating Cybersecurity Capabilities and Risks of Language Models},
  author    = {Zhang, Andy K. and others},
  booktitle = {International Conference on Learning Representations (ICLR)},
  year      = {2025}
}

@article{mao2023gptdriver,
  title   = {{GPT-Driver}: Learning to Drive with {GPT}},
  author  = {Mao, Jiageng and Qian, Yuxi and Ye, Junjie and Zhao, Hang and Wang, Yue},
  journal = {arXiv preprint arXiv:2310.01415},
  year    = {2023}
}

@misc{anthropic2025skills,
  title        = {Introducing Agent Skills},
  author       = {{Anthropic}},
  howpublished = {\url{https://www.anthropic.com/news/skills}},
  year         = {2025},
}

@article{skills2026analysis,
  title={Agent skills: A data-driven analysis of claude skills for extending large language model functionality},
  author={Ling, George and Zhong, Shanshan and Huang, Richard},
  journal={arXiv preprint arXiv:2602.08004},
  year={2026}
}

@misc{anthropic2026claudecode,
  title        = {{Claude Code}},
  author       = {{Anthropic}},
  howpublished = {\url{https://code.claude.com/}},
  year         = {2026},
}

@misc{openai2026codex,
  title        = {{Codex}: {AI} Coding Partner from {OpenAI}},
  author       = {{OpenAI}},
  howpublished = {\url{https://openai.com/codex/}},
  year         = {2026},
}

@misc{snyk2026toxicskills,
  title        = {Snyk Finds Prompt Injection in 36\%, 1467 Malicious Payloads in a {ToxicSkills} Study of Agent Skills Supply Chain Compromise},
  author       = {{Snyk Security Labs}},
  howpublished = {\url{https://snyk.io/blog/toxicskills-malicious-ai-agent-skills-clawhub/}},
  year         = {2026},
}

@misc{koi2026clawhavoc,
  title        = {{ClawHavoc}: 341 Malicious {Clawed} Skills Found by the Bot They Were Targeting},
  author       = {{Koi Security}},
  howpublished = {\url{https://www.koi.ai/blog/clawhavoc-341-malicious-clawedbot-skills-found-by-the-bot-they-were-targeting}},
  year         = {2026},
  note         = {Accessed 2026}
}

@misc{clawhub2026archive,
  title        = {{GitHub -- openclaw/skills: All versions of all skills that are on clawhub.com archived $\cdot$ GitHub}},
  author       = {{OpenClaw}},
  howpublished = {\url{https://github.com/openclaw/skills}},
  note         = {Public archive of all skills published to ClawHub, including malicious skills removed from the live marketplace. Archived snapshot: \url{https://web.archive.org/web/20260406111320/https://github.com/openclaw/skills}},
  year         = {2026},
}

@misc{marketsandmarkets2026agents,
  title        = {{AI} Agents Market Report 2025--2030, by Application, Geo, Tech},
  author       = {{MarketsandMarkets}},
  howpublished = {\url{https://www.marketsandmarkets.com/Market-Reports/ai-agents-market-15761548.html}},
  year         = {2025},
  note         = {Report code TC 9264. Accessed 2026}
}

@misc{scanner_huifer,
  title = {skill-security-scan}, author = {{HuiFer}},
  howpublished = {\url{https://github.com/huifer/skill-security-scan}}, year = {2026}}

@misc{scanner_vigile,
  title = {vigile-scan}, author = {{Vigile.dev}},
  howpublished = {\url{https://github.com/Vigile-ai/vigile-scan}}, year = {2026}}

@misc{scanner_qualixar,
  title = {{SkillFortify}: Formal Capability Verification for Agent Skills}, author = {Bhardwaj, Varun Pratap},
  howpublished = {\url{https://github.com/qualixar/skillfortify}}, year = {2026}}

@misc{scanner_patidarganesh,
  title = {SkillScanner}, author = {Patidar, Ganesh},
  howpublished = {\url{https://github.com/patidarganesh/SkillScanner}}, year = {2026}}

@misc{scanner_caterpillar,
  title = {caterpillar}, author = {{Alice IO}},
  howpublished = {\url{https://github.com/alice-dot-io/caterpillar}}, year = {2026}}

@misc{scanner_cisco,
  title = {Cisco {AI} Defense Skill Scanner}, author = {{Cisco AI Defense}},
  howpublished = {\url{https://github.com/cisco-ai-defense/skill-scanner}}, year = {2026}}

@misc{scanner_headyzhang,
  title = {agent-audit}, author = {{HeadyZhang}},
  howpublished = {\url{https://github.com/HeadyZhang/agent-audit}}, year = {2026}}

@misc{scanner_nova,
  title = {nova-proximity}, author = {Roccia, Thomas},
  howpublished = {\url{https://github.com/fr0gger/proximity}}, year = {2026}}

@misc{scanner_pors,
  title = {skill-audit}, author = {Pors, Mark},
  howpublished = {\url{https://github.com/pors/skill-audit}}, year = {2026}}

@inproceedings{greshake2023ipi,
  title     = {Not What You've Signed Up For: Compromising Real-World {LLM}-Integrated Applications with Indirect Prompt Injection},
  author    = {Greshake, Kai and Abdelnabi, Sahar and Mishra, Shailesh and Endres, Christoph and Holz, Thorsten and Fritz, Mario},
  booktitle = {Proceedings of the 16th ACM Workshop on Artificial Intelligence and Security (AISec)},
  year      = {2023}
}

@inproceedings{zhan2024injecagent,
  title     = {{InjecAgent}: Benchmarking Indirect Prompt Injections in Tool-Integrated Large Language Model Agents},
  author    = {Zhan, Qiusi and Liang, Zhixiang and Ying, Zifan and Kang, Daniel},
  booktitle = {Findings of the Association for Computational Linguistics (ACL)},
  year      = {2024}
}

@inproceedings{debenedetti2024agentdojo,
  title     = {{AgentDojo}: A Dynamic Environment to Evaluate Prompt Injection Attacks and Defenses for {LLM} Agents},
  author    = {Debenedetti, Edoardo and Zhang, Jie and Balunovi\'c, Mislav and Beurer-Kellner, Luca and Fischer, Marc and Tram\`er, Florian},
  booktitle = {Advances in Neural Information Processing Systems (NeurIPS) Datasets and Benchmarks Track},
  year      = {2024}
}

@article{qu2026supplychain,
  title   = {Supply-Chain Poisoning Attacks Against {LLM} Coding Agent Skill Ecosystems},
  author  = {Qu, Yubin and Liu, Yi},
  journal = {arXiv preprint arXiv:2604.03081},
  year    = {2026}
}

@article{skillject2026,
  title   = {{SkillJect}: Effectively Automating Skill-Based Prompt Injection for Skill-Enabled Agents},
  author  = {Jia, Xiaojun and Liao, Jie and Qin, Simeng and Gu, Jindong and Ren, Wenqi and Cao, Xiaochun and Liu, Yang and Torr, Philip},
  journal = {arXiv preprint arXiv:2602.14211},
  year    = {2026}
}

@article{schmotz2026skillinject,
  title   = {{Skill-Inject}: Measuring Agent Vulnerability to Skill File Attacks},
  author  = {Schmotz, David and Beurer-Kellner, Luca and Abdelnabi, Sahar and Andriushchenko, Maksym},
  journal = {arXiv preprint arXiv:2602.20156},
  year    = {2026}
}

@article{liu2026skillswild,
  title   = {Agent Skills in the Wild: An Empirical Study of Security Vulnerabilities at Scale},
  author  = {Liu, Yi and Wang, Weizhe},
  journal = {arXiv preprint arXiv:2601.10338},
  year    = {2026}
}

@article{chen2026credential,
  title   = {How Your Credentials Are Leaked by {LLM} Agent Skills: An Empirical Study},
  author  = {Chen, Zhihao and Zhang, Ying},
  journal = {arXiv preprint arXiv:2604.03070},
  year    = {2026}
}

@article{malskills2026wild,
  title   = {``Do Not Mention This to the User'': Detecting and Understanding Malicious Agent Skills in the Wild},
  author  = {Liu, Yi and Chen, Zhihao and Zhang, Yanjun and Deng, Gelei and Li, Yuekang and Ning, Jianting and Zhang, Leo Yu},
  journal = {arXiv preprint arXiv:2602.06547},
  year    = {2026}
}

@article{malskillbench2026,
  title   = {{MalSkillBench}: A Runtime-Verified Benchmark of Malicious Agent Skills},
  author  = {Guo, Wenbo and Zeng, Wei and Liu, Chengwei and Jia, Xiaojun and Xu, Yijia and Tang, Lei and Fang, Yong and Liu, Yang},
  journal = {arXiv preprint arXiv:2606.07131},
  year    = {2026}
}

@article{skillsbench2026,
  title   = {{SkillsBench}: Benchmarking How Well Agent Skills Work Across Diverse Tasks},
  author  = {Li, Xiangyi and others},
  journal = {arXiv preprint arXiv:2602.12670},
  year    = {2026}
}

@article{phantomskill2026,
  title   = {{PhantomSkill}: Malicious Code Injection in Agent Skill Ecosystems},
  author  = {Lin, Yu-Ting and Yu, Chia-Mu},
  journal = {arXiv preprint arXiv:2606.19191},
  year    = {2026}
}

@inproceedings{ohm2020backstabber,
  title     = {Backstabber's Knife Collection: A Review of Open Source Software
               Supply Chain Attacks},
  author    = {Ohm, Marc and Plate, Henrik and Sykosch, Arnold and Meier, Michael},
  booktitle = {Detection of Intrusions and Malware, and Vulnerability Assessment (DIMVA)},
  year      = {2020}
}

@article{skillsieve2026,
  title   = {{SkillSieve}: A Hierarchical Triage Framework for Detecting Malicious {AI} Agent Skills},
  author  = {Hou, Yinghan and Yang, Zongyou and Pang, Zaihu and Ma, Xiujun},
  journal = {arXiv preprint arXiv:2604.06550},
  year    = {2026}
}

@article{cascade2026,
  title   = {{CASCADE}: A Cascaded Hybrid Defense Architecture for Prompt-Injection Detection in {MCP}-based Systems},
  author  = {{\.I}pek Abas{\i}kele{\c{s}} Turgut and Edip G{\"u}m{\"u}{\c{s}}},
  journal = {arXiv preprint arXiv:2604.17125},
  year    = {2026}
}

@article{malskills2026,
  title   = {"Elementary, My Dear Watson." Detecting Malicious Skills via Neuro-Symbolic Reasoning across Heterogeneous Artifacts},
  author  = {Wang, Shenao and He, Junjie},
  journal = {arXiv preprint arXiv:2603.27204},
  year    = {2026}
}

@inproceedings{shi2024judgedeceiver,
  title     = {Optimization-based Prompt Injection Attack to {LLM}-as-a-Judge},
  author    = {Shi, Jiawen and Yuan, Zenghui and Liu, Yinuo and Huang, Yue and Zhou, Pan and Sun, Lichao and Gong, Neil Zhenqiang},
  booktitle = {Proceedings of the ACM SIGSAC Conference on Computer and Communications Security (CCS)},
  year      = {2024}
}

@inproceedings{raina2024judge,
  title     = {Is {LLM}-as-a-Judge Robust? Investigating Universal Adversarial Attacks on Zero-shot {LLM} Assessment},
  author    = {Raina, Vyas and Liusie, Adian and Gales, Mark},
  booktitle = {Proceedings of the Conference on Empirical Methods in Natural Language Processing (EMNLP)},
  year      = {2024}
}

@article{nasr2025attacker,
  title   = {The Attacker Moves Second: Stronger Adaptive Attacks Bypass Defenses Against {LLM} Jailbreaks and Prompt Injections},
  author  = {Nasr, Milad and Carlini, Nicholas and others},
  journal = {arXiv preprint arXiv:2510.09023},
  year    = {2025}
}

@inproceedings{hackett2025bypassing,
  title     = {Bypassing {LLM} Guardrails: An Empirical Analysis of Evasion Attacks against Prompt Injection and Jailbreak Detection Systems},
  author    = {Hackett, William and others},
  booktitle = {ACL Workshop on Large Language Model Security (LLMSEC)},
  year      = {2025}
}

@article{cohen1987virus,
  title   = {Computer Viruses: Theory and Experiments},
  author  = {Cohen, Fred},
  journal = {Computers \& Security},
  volume  = {6},
  number  = {1},
  pages   = {22--35},
  year    = {1987}
}

@inproceedings{chess2000undetectable,
  title     = {An Undetectable Computer Virus},
  author    = {Chess, David M. and White, Steve R.},
  booktitle = {Proceedings of the Virus Bulletin Conference},
  year      = {2000}
}

@inproceedings{christodorescu2005semanticsaware,
  title     = {Semantics-Aware Malware Detection},
  author    = {Christodorescu, Mihai and Jha, Somesh and Seshia, Sanjit A. and Song, Dawn and Bryant, Randal E.},
  booktitle = {IEEE Symposium on Security and Privacy (S\&P)},
  year      = {2005}
}

@inproceedings{moser2007limits,
  title     = {Limits of Static Analysis for Malware Detection},
  author    = {Moser, Andreas and Kruegel, Christopher and Kirda, Engin},
  booktitle = {23rd Annual Computer Security Applications Conference (ACSAC)},
  year      = {2007}
}

@inproceedings{lindorfer2011environmentsensitive,
  title     = {Detecting Environment-Sensitive Malware},
  author    = {Lindorfer, Martina and Kolbitsch, Clemens and Comparetti, Paolo Milani},
  booktitle = {International Symposium on Recent Advances in Intrusion Detection (RAID)},
  year      = {2011}
}

@inproceedings{kirat2014barecloud,
  title     = {{BareCloud}: Bare-metal Analysis-based Evasive Malware Detection},
  author    = {Kirat, Dhilung and Vigna, Giovanni and Kruegel, Christopher},
  booktitle = {USENIX Security Symposium},
  pages     = {287--301},
  year      = {2014}
}

@inproceedings{peng2014xforce,
  title     = {{X-Force}: Force-Executing Binary Programs for Security Applications},
  author    = {Peng, Fei and Deng, Zhui and Zhang, Xiangyu and Xu, Dongyan and Lin, Zhiqiang and Su, Zhendong},
  booktitle = {USENIX Security Symposium},
  pages     = {829--844},
  year      = {2014}
}

@article{borello2008metamorphic,
  title   = {Code Obfuscation Techniques for Metamorphic Viruses},
  author  = {Borello, Jean-Marie and M\'e, Ludovic},
  journal = {Journal in Computer Virology},
  volume  = {4},
  number  = {3},
  pages   = {211--220},
  year    = {2008}
}

@article{anderson2018evasion,
  title   = {Learning to Evade Static {PE} Machine Learning Malware Models via Reinforcement Learning},
  author  = {Anderson, Hyrum S. and Kharkar, Anant and Filar, Bobby and Evans, David and Roth, Phil},
  journal = {arXiv preprint arXiv:1801.08917},
  year    = {2018}
}

@inproceedings{pierazzi2020problemspace,
  title     = {Intriguing Properties of Adversarial {ML} Attacks in the Problem Space},
  author    = {Pierazzi, Fabio and Pendlebury, Feargus and Cortellazzi, Jacopo and Cavallaro, Lorenzo},
  booktitle = {IEEE Symposium on Security and Privacy (S\&P)},
  year      = {2020}
}

@inproceedings{newsome2005dynamic,
  title     = {Dynamic Taint Analysis for Automatic Detection, Analysis, and Signature Generation of Exploits on Commodity Software},
  author    = {Newsome, James and Song, Dawn},
  booktitle = {Network and Distributed System Security Symposium (NDSS)},
  year      = {2005}
}

@inproceedings{clause2007dytan,
  title     = {Dytan: A Generic Dynamic Taint Analysis Framework},
  author    = {Clause, James and Li, Wanchun and Orso, Alessandro},
  booktitle = {International Symposium on Software Testing and Analysis (ISSTA)},
  year      = {2007}
}

@inproceedings{kemerlis2012libdft,
  title     = {libdft: Practical Dynamic Data Flow Tracking for Commodity Systems},
  author    = {Kemerlis, Vasileios P. and Portokalidis, Georgios and Jee, Kangkook and Keromytis, Angelos D.},
  booktitle = {ACM SIGPLAN/SIGOPS International Conference on Virtual Execution Environments (VEE)},
  year      = {2012}
}

@inproceedings{yin2007panorama,
  title     = {Panorama: Capturing System-wide Information Flow for Malware Detection and Analysis},
  author    = {Yin, Heng and Song, Dawn and Egele, Manuel and Kruegel, Christopher and Kirda, Engin},
  booktitle = {ACM Conference on Computer and Communications Security (CCS)},
  year      = {2007}
}

@inproceedings{enck2010taintdroid,
  title     = {{TaintDroid}: An Information-Flow Tracking System for Realtime Privacy Monitoring on Smartphones},
  author    = {Enck, William and Gilbert, Peter and Chun, Byung-Gon and Cox, Landon P. and Jung, Jaeyeon and McDaniel, Patrick and Sheth, Anmol N.},
  booktitle = {USENIX Symposium on Operating Systems Design and Implementation (OSDI)},
  year      = {2010}
}

@inproceedings{schwartz2010all,
  title     = {All You Ever Wanted to Know About Dynamic Taint Analysis and Forward Symbolic Execution (but Might Have Been Afraid to Ask)},
  author    = {Schwartz, Edward J. and Avgerinos, Thanassis and Brumley, David},
  booktitle = {IEEE Symposium on Security and Privacy (S\&P)},
  year      = {2010}
}

@inproceedings{wu2025isolategpt,
  title     = {{IsolateGPT}: An Execution Isolation Architecture for {LLM}-Based Agentic Systems},
  author    = {Wu, Yuhao and Roesner, Franziska and Kohno, Tadayoshi and Zhang, Ning and Iqbal, Umar},
  booktitle = {Network and Distributed System Security Symposium (NDSS)},
  year      = {2025}
}

@article{shi2025progent,
  title   = {Progent: Securing {AI} Agents with Privilege Control},
  author  = {Shi, Tianneng and others},
  journal = {arXiv preprint arXiv:2504.11703},
  year    = {2025}
}

@inproceedings{wang2026agentspec,
  title     = {{AgentSpec}: Customizable Runtime Enforcement for Safe and Reliable {LLM} Agents},
  author    = {Wang, Haoyu and Poskitt, Christopher M. and Sun, Jun},
  booktitle = {Proceedings of the 48th International Conference on Software Engineering (ICSE)},
  year      = {2026}
}

@inproceedings{tsai2025conseca,
  title     = {Contextual Agent Security: A Policy for Every Purpose},
  author    = {Tsai, Lillian and others},
  booktitle = {Workshop on Hot Topics in Operating Systems (HotOS)},
  year      = {2025}
}

@article{debenedetti2025camel,
  title   = {Defeating Prompt Injections by Design},
  author  = {Debenedetti, Edoardo and Shumailov, Ilia and Fan, Tianqi and Hayes, Jamie and Carlini, Nicholas and Fabian, Daniel and Kern, Christoph and Shi, Chongyang and Terzis, Andreas and Tram\`er, Florian},
  journal = {arXiv preprint arXiv:2503.18813},
  year    = {2025}
}

@article{zhong2025rtbas,
  title   = {{RTBAS}: Defending {LLM} Agents Against Prompt Injection and Privacy Leakage},
  author  = {Zhong, Peter and others},
  journal = {arXiv preprint arXiv:2502.08966},
  year    = {2025}
}

@article{costa2025fides,
  title   = {Securing {AI} Agents with Information-Flow Control},
  author  = {Costa, Manuel and K\"opf, Boris and others},
  journal = {arXiv preprint arXiv:2505.23643},
  year    = {2025}
}

@inproceedings{forrest1996sense,
  title     = {A Sense of Self for Unix Processes},
  author    = {Forrest, Stephanie and Hofmeyr, Steven A. and Somayaji, Anil and Longstaff, Thomas A.},
  booktitle = {IEEE Symposium on Security and Privacy (S\&P)},
  year      = {1996}
}

@article{hofmeyr1998intrusion,
  title   = {Intrusion Detection Using Sequences of System Calls},
  author  = {Hofmeyr, Steven A. and Forrest, Stephanie and Somayaji, Anil},
  journal = {Journal of Computer Security},
  volume  = {6},
  number  = {3},
  pages   = {151--180},
  year    = {1998}
}

@inproceedings{sekar2024eaudit,
  title     = {{eAudit}: A Fast, Scalable and Deployable Audit Data Collection System},
  author    = {Sekar, R. and others},
  booktitle = {IEEE Symposium on Security and Privacy (S\&P)},
  year      = {2024}
}

@article{ji2026taming,
  title   = {Taming Various Privilege Escalation in {LLM}-Based Agent Systems: A Mandatory Access Control Framework},
  author  = {Ji, Zimo and Wu, Daoyuan and Jiang, Wenyuan and Ma, Pingchuan and Li, Zongjie and Gao, Yudong and Wang, Shuai and Li, Yingjiu},
  journal = {arXiv preprint arXiv:2601.11893},
  year    = {2026}
}

@article{ji2025taxonomy,
  title   = {Taxonomy, Evaluation and Exploitation of {IPI}-Centric {LLM} Agent Defense Frameworks},
  author  = {Ji, Zimo and Wang, Xunguang and Li, Zongjie and Ma, Pingchuan and Gao, Yudong and Wu, Daoyuan and Yan, Xincheng and Tian, Tian and Wang, Shuai},
  journal = {arXiv preprint arXiv:2511.15203},
  year    = {2025}
}

\end{document}